\documentclass[a4paper,11pt]{article}
\usepackage{graphicx} 
\graphicspath{{Images/}}
\usepackage[export]{adjustbox}
\usepackage{multirow}
\usepackage{hyperref}

\title{A Systematic Mapping Study on SDN Controllers for Enhancing Security in IoT Networks}

\author{
Charles Oredola, Adnan Ashraf \\
Faculty of Science and Engineering \\
\AA bo Akademi University, Turku, Finland \\
\texttt{charles.oredola@abo.fi, adnan.ashraf@abo.fi}
}
\date{}

\def\BibTeX{{\rm B\kern-.05em{\sc i\kern-.025em b}\kern-.08em
    T\kern-.1667em\lower.7ex\hbox{E}\kern-.125emX}}

\begin{document}

\maketitle
\begin{abstract}

 \textit{Context:} The increase in Internet of Things (IoT) devices gives rise to an increase in deceptive manipulations by malicious actors. These actors should be prevented from targeting the IoT networks. Cybersecurity threats have evolved and become dynamically sophisticated, such that they could exploit any vulnerability found in IoT networks. However, with the introduction of the Software Defined Network (SDN) in the IoT networks as the central monitoring unit, IoT networks are less vulnerable and less prone to threats. 

\noindent\textit{Objective:} To present a comprehensive and unbiased overview of the state-of-the-art on IoT networks security enhancement using SDN controllers.

\noindent\textit{Method:} We review the current body of knowledge on enhancing the security of IoT networks using SDN with a Systematic Mapping Study (SMS) following the established guidelines.

\noindent\textit{Results:} The SMS result comprises 33 primary studies analyzed against four major research questions. The SMS highlights current research trends and identifies gaps in the SDN-IoT network security.

\noindent\textit{Conclusion:} We conclude that the SDN controller architecture commonly used for securing IoT networks is the centralized controller architecture. However, this architecture is not without its limitations. Additionally, the predominant technique utilized for risk mitigation is machine learning.


\end{abstract}

\section{Introduction}
\label{section:Intro}

Several preventive approaches \cite{47} have been developed to counter the effects of malicious acts in the Internet of Things (IoT) devices and networks in the past decades. There are a number of attacks targeted at IoT devices, several of which are flooding attacks such as Denial of Service (DoS) and Distributed Denial of Service (DDoS). It has been forecasted that by 2025, several billion IoT devices will be in use \cite{S8, S17}. As IoT devices increase in number, so do the threats that come with the increment. In recent years, the Software-Defined Networking (SDN) \cite{S1, S3} paradigm has gained momentum in network security with its effective network management features. With the introduction of SDN in IoT networks, there is an assurance of a more secure network. SDN has proved to be a promising technology to deal with the security challenges in IoT networks by providing centralized network management, programmability, and automation. SDN separates the control plane from the data plane, enabling dynamic network configuration and policy enforcement to enhance security and mitigate threats in IoT deployments. However, SDN controllers can be vulnerable. This can allow threat actors to infiltrate the IoT network through the SDN interface \cite{S19, S28}. Therefore, there is a need to secure the SDN interface itself. Approaches like machine learning algorithms \cite{S8, S10, S14} IP packet filtering \cite{S16}, trust management algorithms \cite{S9}, policy-based techniques \cite{S15}, and smart contract-based frameworks \cite{S24} are some of the preventive techniques proposed by scientists for securing SDN-based IoT networks. These approaches have been used with different SDN controller architectures to detect and prevent intrusions in IoT networks.

Despite the growing interest in leveraging SDN for IoT security, there is a lack of comprehensive research on the various SDN controller architectures for enhancing security in IoT networks. Therefore, this Systematic Mapping Study (SMS) aims to provide an overview of the existing literature on SDN controller architectures for IoT security and identify trends, gaps, and research directions. We formulate our selection criteria, create a classification scheme, and present four main research questions on IoT security using SDN. We have selected 33 primary studies and analyzed them against the research questions. We also highlight various drawbacks and identify current trends and research gaps in SDN-based IoT network security.

The rest of the paper is structured as follows. We provide some background knowledge of SDN architectures and describe the IoT Architecture in Section~\ref{section:SDN}. Section~\ref{section:design} outlines the design of our SMS and the main steps involved. The results of the study are presented in Section~\ref{section:results}. We address threats to the validity of this SMS in Section~\ref{section:threats}. Section~\ref{section:conclusions} presents our conclusions.

\section{SDN Controller Architecture and IoT}
\label{section:SDN}

This section briefly describes the SDN controller architecture. It also provides an overview of the IoT architecture and IoT network security.

\subsection{SDN Planes Classification}
\label{subsection:SDN Planes}

SDN can be classified into three planes: the control plane, the data plane, and the application plane. The control plane is responsible for network management and routing decisions. The data plane transports network data packets according to control plane directives. The application plane contains network policy implementations, security configurations, and management functions.

The three planes in SDN architecture communicate with one another via the northbound interface and southbound interface. Communication between the SDN controller and the data plane is enabled through protocols like OpenFlow, while communication with the application plane is facilitated through various APIs and interfaces \cite{34}.

\subsection{SDN Controller Configuration}
\label{subsection:SDN configuration}

Globally, the SDN is a centralized logical network device, positioned at the center of the entire network to oversee and manage all events within the network. However, the SDN controller can likewise be decentralized in a network. The SDN controller decentralization can be divided into two types according to the network's configuration: Core-Decentralization and Distributed-Decentralization.

\subsubsection{Centralized SDN Controller} In this configuration, a single SDN controller is located at the core of the network \cite{S3, S7, S8, S9, S10}. This controller is responsible for managing the entire network, making routing decisions, and overseeing all network events. While this setup simplifies network management, it is vulnerable to various types of attacks  \cite{S19}, including Man-in-the-middle (MitM) attacks.

\subsubsection{Core-Decentralized SDN Controller} In this configuration, each node in the network is empowered to make routing decisions independently.
Decentralized decision-making at the node level enhances network resilience and reduces reliance on a central controller \cite{S17, S24, S25, S33}.
Nodes can adapt dynamically to network conditions, improving overall network performance and robustness.

\subsubsection{Distributed-Decentralized SDN Controller} This configuration combines elements of both centralized and decentralized control. Each node in the network has its own SDN controller capable of making local decisions. However, these controllers also report to a central SDN controller which makes overarching decisions for the network \cite{S2, S11, S27}. This hybrid approach provides a balance between local autonomy and centralized oversight, offering flexibility and scalability.

\subsection{IoT Architecture Overview}
\label{subsection:IoT Overview}
IoT architecture refers to the framework of interconnected devices, systems, protocols, and components that enable the collection, processing, and exchange of data in IoT ecosystems. It provides a structured approach to designing and implementing IoT solutions. An overview of the key components and layers in the IoT architecture is as follows.

\subsubsection{Perception Layer} The perception layer comprises IoT devices that interact with the physical environment using sensors, actuators, and other embedded technologies.

\subsubsection{Network Layer} Facilitates communication between IoT devices, gateways, and back-end systems via wired or wireless networks \cite{45}.

\subsubsection{Middleware Layer} Provides services for managing, processing, and integrating IoT data into higher-level applications and systems \cite{44}.

\subsubsection{Application Layer} Utilizes IoT data to deliver value-added functionality and insights through applications such as smart home systems, industrial automation, asset tracking, and healthcare monitoring \cite{46}.

\subsubsection{Security Layer} Ensures the confidentiality, integrity, and availability of IoT data and devices, employing authentication, authorization, encryption, and secure communication protocols. Implements intrusion detection/prevention systems (IDS/IPS) \cite{47} and well-structured Identity and Access Management (IAM) to protect against unauthorized access, data breaches, and other security threats.

\subsubsection{Management and Orchestration Layer} Responsible for provisioning, configuring, monitoring, and managing IoT devices, networks, and applications.

As discussed in this section, the existing SDN architectures for IoT networks consider some cybersecurity issues. However, a comprehensive study of cybersecurity issues in SDN-based IoT networks is needed to understand the issues and enhance the security of SDN-based IoT networks.

\section{Study Design}
\label{section:design}

\begin{figure}[h!]
\includegraphics[scale=0.347]{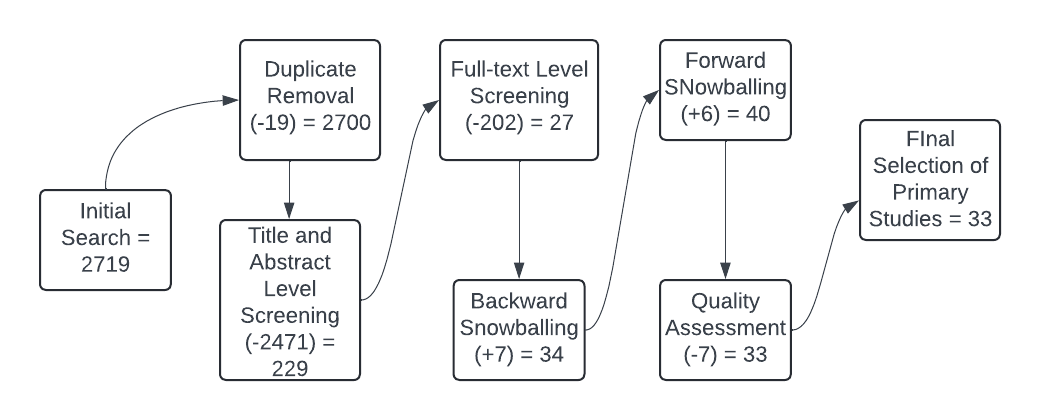}
\centering
\caption{The SMS Process Flow.}
\label{fig:Flow}
\end{figure}

This section presents the review method adopted following the established guidelines for conducting an SMS in \cite{37}. We conducted a thorough search of some specific digital libraries (DLs) by adopting the protocols for developing an SMS. These protocols are as follows; We collected the most relevant primary studies by specifying the appropriate search strings. We analyzed the abstracts of the primary studies collected from the considered databases. We removed the duplicates and filtered out the irrelevant primary studies. We filtered the primary studies further by doing a full-text level screening. We performed snowball sampling~\cite{38} to find additional relevant primary studies. We excluded some studies that did not meet the minimum quality assessment criteria during the quality assessment evaluation. We used the filtered primary studies extracted from the quality assessment evaluation stage to answer the research questions. Figure \ref{fig:Flow} shows the number of included and excluded primary studies at each stage.

\subsection{Research questions}
\label{subsection:Research questions}

The research questions were focused on identifying the research gaps in this research area by classifying the selected studies on finding a secure method for SDN-based IoT networks. The following are the research questions:

\noindent \textbf{RQ1:} What kind of security issues exist in SDN-based IoT networks (attacks, threats, anomalies, vulnerabilities)?

\noindent \textbf{RQ2:} What approaches are used to address security issues in SDN-based IoT networks?

\noindent \textbf{RQ2.1:} What methods/techniques/algorithms are used to address security issues in SDN-based IoT networks?

\noindent \textbf{RQ2.2:}     Which SDN controller architectures are used to address security issues in SDN-based IoT networks?

\noindent \textbf{RQ2.3:} What kind of data analysis techniques can be used to address security issues in SDN-based IoT networks?

\noindent \textbf{RQ3:} What kind of SDN datasets are available to address security issues in SDN-based IoT networks?

\noindent \textbf{RQ4:} What simulation tools and techniques are used for intrusion detection and prevention in SDN-based IoT networks?

\subsection{Source selection}
\label{subsection:Source selection}

\begin{figure}[h!]
\includegraphics[scale=0.52]{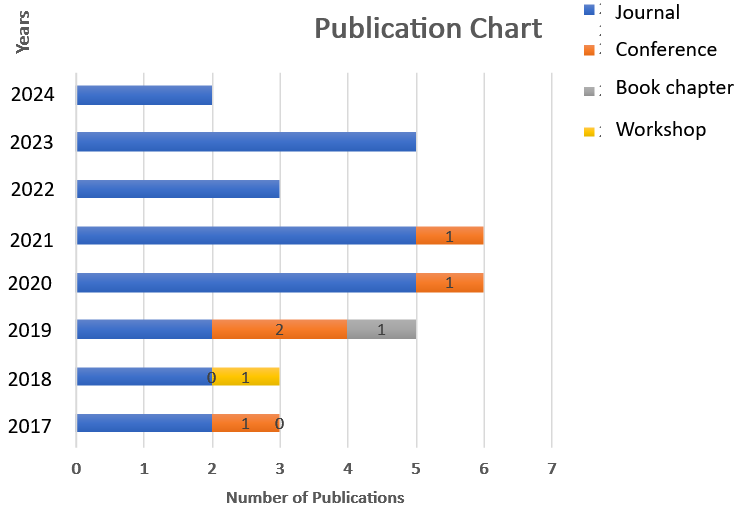}
\centering
\caption{The distribution of the publication types per number of publications considered in the selected primary studies.}
\label{fig:publication chart}
\end{figure}

Using the formulated search strings, we carried out an intensive search of the primary studies. We considered five DLs for conducting our systematic search; \textit{IEEE Xplore, ACM, Science Direct, Wiley,} and \textit{Scopus}. These electronic libraries contain important scientific journals, workshops, book chapters, and conference publications in the relevant fields. We considered recent publications in our search. The publication distribution can be seen in Figure \ref{fig:publication chart}.

\subsection{Search string}
\label{subsection:Search string}

The search string we used to collect our initial primary studies is presented in Table \ref{table:searchstring}.

\begin{table}[h!]
\caption{Search String}
\label{table:searchstring}
\centering
\begin{tabular}{|p{11.5cm}|}
 \hline
  {(SDN OR “Software-defined network”) AND (IoT OR “Internet of things”) AND (Secur* OR Vulnerabilit* OR Anomal* OR Attack* OR Threat*) AND (Method* OR Technique* OR Algorithm* OR Architecture)}\\
   \hline
\end{tabular}
\end{table}

The following fields are the database fields that were used during the search for primary studies: title, abstract, and keywords. We used wildcard operators to accommodate different variations of the terms, for instance, we used secur* for security, etc. Table \ref{table:Databases} lists the number of primary studies found from each database.

\begin{table}[h!]
\caption{Initial Databases Search}
\label{table:Databases}
\centering
\begin{tabular}{|p{4cm}|c|}
 \hline
 \textbf{Database} & \textbf{Results} \\
 \hline
 IEEEXplore & 579 \\
 \hline
 ACM & 652 \\
 \hline
 Science Direct & 546 \\
 \hline
 Scopus & 584 \\
 \hline
 Wiley & 358 \\
 \hline
 \textbf{Total} & \textbf{2719} \\
 \hline
\end{tabular}
\end{table}

\subsection{Study selection criteria}
\label{subsection:Study Selection}

To select the most relevant primary studies for our SMS, we considered only peer-reviewed and published research papers with adequate technical information. The study selection was carried out by two researchers. After querying the databases for the initial primary studies, the search results were filtered to get the most technically informative research papers. Snowballing (both forward and backward snowballing) \cite{38} activities were done to complement additional electronic searches.
The two researchers met to discuss the inclusion and exclusion criteria to remove researcher's bias. Both researchers agreed on the following criteria for both inclusion and exclusion:

\textbf{Inclusion criteria:}
\begin{itemize}
    \item All research papers with the search string as discussed in subsection~\ref{subsection:Search string} in their abstracts, titles, and keywords are included
    \item Written in English
    \item Published in a peer-reviewed journal, book chapter, conference, or workshop of computer science, computer engineering, or software engineering.
    \item In addition, if several papers present the same approach, only the most recent will be included.
\end{itemize}

\textbf{Exclusion criteria:} The Exclusion criteria is the direct negation of the Inclusion Criteria.


\subsection{Study quality assessment}
\label{subsection:Study quality}

We evaluated the quality of our selected primary studies against some set of already designed research questions (as suggested in \cite{41}) that suit our research purpose. The quality assessment criteria questions include the following:
\begin{enumerate}
    \item Are the security challenges and mitigations in SDN-based IoT networks clearly defined?
	\item Are the attacks, vulnerabilities and anomalies in SDN-IoT discussed in detail?
	\item Are the approaches to reduce SDN vulnerabilities and anomalies clearly stated?
    \item Is a simulation or a prototype implementation presented?
	\item Is the experimental design and setup clearly described?
	\item Is the experimental result clearly analyzed?
	\item Are the results from different experiments included?
	\item Are the experimental results compared with other state-of-the-art approaches?
\end{enumerate}

For each question in the questionnaires above, a three-level, numeric scale was used \cite{40} to evaluate the selected primary studies. The levels are:
\begin{itemize}
    \item Answers a question (2 points),
    \item Does not answer question (0 point), and
    \item Partially answers a question (1 point).
\end{itemize}

Based on the questionnaires and the numeric scale, a study can score a maximum of 16 and a minimum of 0 points. Moreover, for each study, the independent scores from the two researchers were aggregated by computing the arithmetic mean \cite{39}. Our considered score for merit was 8. Therefore, a study that scored less than 8/16 points was excluded in the quality assessment stage due to its lack of quality merit.

\subsection{Data extraction}
\label{subsection:Data extraction}

We extracted the information relevant to our research from the selected primary studies by creating a free form for the data extraction stage. We gathered two types of information from each study: general information, which includes the paper title, the authors, and the publication year; and information relevant to the research questions.
During the data extraction stage, there were consensus meetings and discussions to settle all discrepancies and disagreements to mitigate errors and rule out all researcher's bias \cite{40}. This was done by allocating the data extraction and data validation tasks to different researchers.



\subsection{Schedule of the study}
\label{subsection:Schedule study}


The important aspects of the study schedule, which include the search for primary studies, backward snowball sampling, and forward snowball sampling had the deadline dates 19 December 2023, 19 February 2024, and 22 February 2024, respectively. The phases in our mapping procedure seem to be in order, in reality, several of them were iterative in our experience. For example, going through the selected primary studies, the data extraction phase updates throughout the process as new information is discovered, requiring more time and effort than expected. Additionally, we discovered that it took a lot of time and effort to present and arrange the mapping results completely and logically. It should be noted that the search for primary studies was concluded on 19 December 2023. As such, the findings do not include any publications published after 19 December 2023.


\section{Results of the Study}
\label{section:results}

In this section, we analyze the answers to the four major research questions listed in subsection~\ref{subsection:Research questions} with their sub-questions according to our study.

\subsection {RQ1: What kind of security issues exist in SDN-based IoT networks (attacks, threats, anomalies, vulnerabilities)?}

The common types of attacks in the peer-reviewed journals and other publications considered are DDoS, DoS, Man-in-the-middle (MiTM), Zero-day attacks, New-flow attacks, and unspecified attacks. DDoS is the most common attack type among IoT networks, this fact can be established among the selected primary studies in Table \ref{table:Security Attacks}, with DDOS being 58\% of the total attack types in the publications considered.


\begin{table}[h!]
\caption{Security Attacks/ Anomalies/ Vulnerabilities types in SDN-based IoT Networks}
\label{table:Security Attacks}
\centering
\scalebox{0.8}{
\begin{tabular}{|l|p{9.5cm}|c|}
 \hline
Security Attacks & Study References & Count \\
 \hline
 DoS	& \cite{S2, S3, S4, S7, S9, S10, S13, S17, S18, S19, S28} & 11 \\
 \hline
 DDoS & \cite{S1, S2, S3, S4, S7, S8, S9, S10, S12, S14, S15, S17, S18, S19, S20, S21, S22, S23, S24, S25, S27, S29, S30, S31, S32} & 25\\
 \hline
 MitM & \cite{S16}	& 1\\
 \hline
 Zero Day Attack & \cite{S32} & 1\\
 \hline
 New Flow Attack	& \cite{S6} & 1\\
 \hline
 Unspecified Attacks & \cite{S11, S26, S33} & 3\\
 \hline
\end{tabular}}
\end{table}

\subsubsection{DoS (Denial of Service)} as the name denotes, is a form of attack targeted at IoT devices in a network by malicious actors to make it unavailable to its intended users by interrupting the normal device functioning capability. This type of attack occurs when a system floods the targeted device in the network with superfluous requests in an attempt to overload the device, thus preventing legitimate requests to the device.

\subsubsection{DDoS (Distributed Denial of Service)} This type of attack is similar to the DoS attack, but It occurs when multiple systems flood the bandwidth of the targeted device. It is carried out from many different sources (from many dummy devices). This is a sophisticated type of attack that cannot be stopped from one device because it comes in from many systems. Therefore, mere ingress filtering may not stop DDoS since the incoming flooding traffic originates from multiple systems from different sources. The common types of DDoS are; Domain Name System (DNS) amplification, Synchronization (SYN) flooding, and User Datagram Protocol (UDP) flooding.

\subsubsection{MiTM} also called the In-path attack is a type of cyber-attack where the attackers secretly insert themselves between two genuine communicating parties to relay and possibly alter the communication. MitM could result in critical or life-threatening issues like unauthorized data access, financial fraud, identity theft, and compromised network security \cite{S16}.

\subsubsection{Zero Day Attack} also called a 0-day attack, is a severe threat that exploits zero-day vulnerabilities \cite{S1, S32} in systems. As the name implies, the device or software vendor has zero-day to fix this kind of threat, this is actually because the attack has already taken place. In this type of attack, the attacker takes advantage of a security flaw in the software, the hardware, or the firmware of a device, as the flaw is either unaddressed or the vendor is unaware of the flaw.

\subsubsection{New Flow Attacks} This can be any type of attack. It can be DoS-based New Flow attack, it can be DDoS-based New Flow Attack \cite{S6, 43}. In this type of attack, the SDN is unaware of the type of packets in-flow, malicious or legitimate, it creates a routing path even if the packets are malicious. New flow attackers aim to overwhelm the SDN-capable switches with unmatched packets, these unmatched packets are regarded as New flows in SDN. To forward the legitimate new flows, or install flow entries, legitimate packets are encapsulated in Packet-In messages and transmitted to the controller. The controller then computes the routing paths and delivers controller-to-switch messages to each switch in the paths. To mitigate against this type of attack, new flow rules are created, and each packet in-flow is inspected against the flow rule.

\subsubsection{Unspecified Attacks} Some authors in the selected primary studies failed to specify the type of attacks they treated. These attacks treated in these papers are therefore classified as unspecified attacks \cite{S11, S26, S33}.

\subsection{RQ2: What approaches are used to address security issues in SDN-based IoT networks?:}

In this subsection, we considered the many approaches employed by the researchers in the selected primary studies to mitigate the common threats in SDN-secured IoT networks. These approaches are the methods/ algorithms/ techniques, the SDN architecture types, and the data analysis method deployed to mitigate the threats.

\subsubsection{RQ2.1 - What methods/techniques/algorithms are used to address security issues in SDN-based IoT networks?}

\begin{figure}[h!]
\includegraphics[scale=0.4]{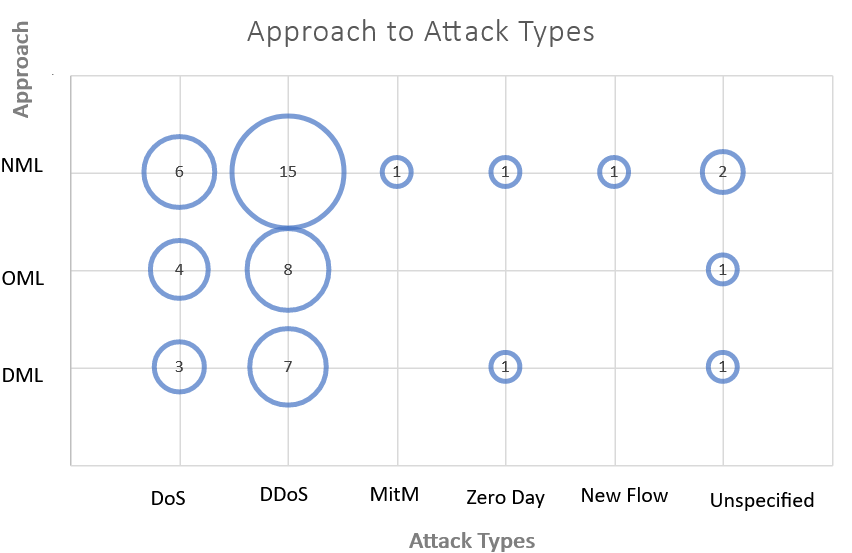}
\centering
\caption{The distribution of the attack types to the mitigation approach}
\label{fig:The Approach Chart}
\end{figure}

Several methods and algorithms have been used by researchers in the past to detect and prevent threats in IoT networks. We have grouped the approaches according to their techniques. The methods mostly considered by researchers for threat mitigation in the past are machine learning algorithms. Although, some researchers used non-machine learning algorithms as seen in Table \ref{table:Proposed Approach} and Figure \ref{fig:The Approach Chart}, machine learning algorithmic approaches for threat mitigation are increasingly gaining popularity in the cyber-world. The number of studies with machine learning approaches outweighs the number of studies with non-machine learning approaches.

\begin{table}[ht!]
\centering
\caption{Proposed Approach}
\label{table:Proposed Approach}
\scalebox{0.8}{
\begin{tabular}{|l|p{3.0cm}|p{5.8cm}|c|}
\hline
    Approach & Algorithm Name & Study References & Count \\
    \hline
    \multirow{2}{*}{Machine Learning} & Deep machine learning models & \cite{S10, S22, S5, S19, S14, S12, S13, S32, S26, S20} & 12  \\ \cline{2-4}

    & Ordinary machine learning models & \cite{S2, S18, S21, S19, S8, S10, S22, S7, S26, S31} & 22 \\
    \hline
    \multicolumn{2}{|l|}{Non-Machine Learning}  & \cite{S1, S3, S4, S6, S7, S9, S11, S14, S15, S16, S17, S19, S23, S24, S25, S27, S28, S29, S30, S32, S33} & 21 \\
\hline
\end{tabular}}
\end{table}

\begin{figure}[b!]
\includegraphics[scale=0.52]{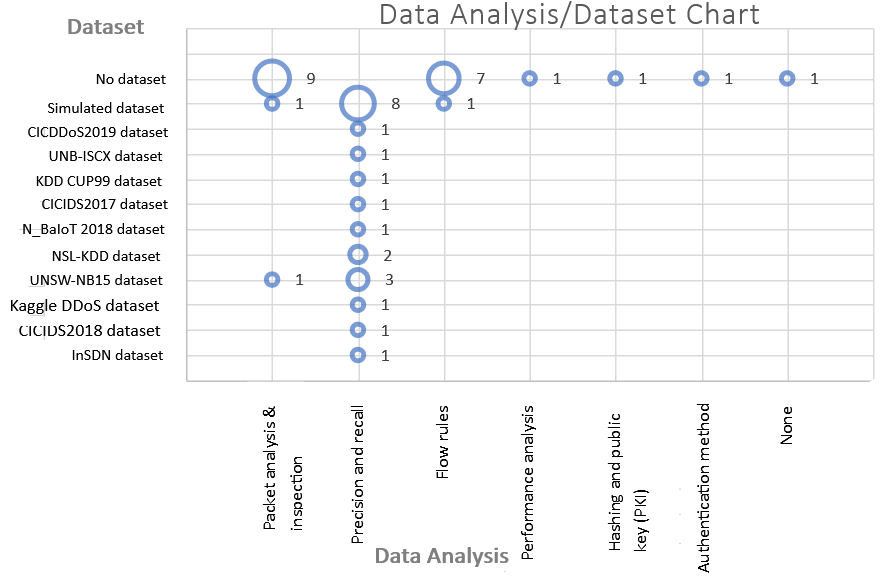}
\centering
\caption{The distribution of the data analysis techniques used to the dataset types considered in the selected primary studies.}
\label{fig:Dataset Distribution}
\end{figure}

 The machine learning algorithms used in the primary studies are further classified into deep learning and ordinary machine learning algorithms. Non-machine learning algorithmic solutions e.g. packet filtering, flow rule, and packet analysis were considered in some primary studies. Each of these papers used at least one of the algorithms as seen in Table \ref{table:Proposed Approach}. A research paper may consider using embedded machine learning algorithms which may include only Deep Machine Learning (DML) models \cite{S12, S13, S32, S10, S22, S26}, Ordinary Machine Learning (OML) models \cite{S7, S8, S18, S19, S21, S22}, both DML models and OML Models \cite{S10, S22, S26} or Machine Learning (ML) algorithm with Non-Machine Learning algorithm (NML) \cite{S7, S32}. The number of DML models that were used is 5 and the number of times that the DML models were used in the selected studies is 12. The number of OML models that were used is 15 and the number of times that the OML models were used in the primary studies is 22. This makes the total number of ML algorithms used in the primary studies 20 and the total number of times when ML algorithms were used 34, lastly, the number of NML techniques that were used is 21, and each of these algorithms was used in at least one study. This gives us a concise conclusion that the common approach used for threat mitigation in IoT networks is mostly done by ML algorithms. The studies that were analyzed against no dataset were analyzed by either packet inspection method, flow rule method, or other methods as specified in Figure \ref{fig:Dataset Distribution}.

 \subsubsection{RQ2.2 - Which SDN controller architectures are used to address security issues in SDN-based IoT networks?}

Subsection~\ref{subsection:SDN configuration} explains the SDN architectural configuration in detail. Considering the selected primary studies, the centralized SDN controller architecture is the most opted for by the researchers. A total of 26 out of the 33 selected primary studies considered using a centralized SDN controller architecture. A single-point-of-failure could be a major disadvantage of networks with this kind of architecture, when the SDN itself is compromised, the entire network could be attacked. However, 4 studies considered using the decentralized architecture, while the remaining 3 studies considered using the distributed controller architecture. Table \ref{table:controller architecture} shows the distribution of the SDN controller architecture usage by the selected studies.

\begin{table}[h!]
\caption{SDN Controller Architecture Usage Distribution}
\label{table:controller architecture}
\centering
\begin{tabular}{|p{2.6cm}|p{7cm}|c|}
 \hline
 SDN Controller Architecture & Study References & Count \\
 \hline
Centralized	& \cite{S1, S3, S4, S5, S6, S7, S8, S9, S10, S12, S13, S14, S15, S16, S18, S20, S21, S22, S23, S26, S28, S29, S30, S31, S32} & 26 \\ \hline
Decentralized & \cite{S17, S24, S25, S33}	& 4\\ \hline
Distributed	& \cite{S2, S11, S27} & 3\\ \hline
\end{tabular}
\end{table}

\subsubsection{RQ2.3 - What kind of data analysis can be used to address security issues in SDN-based IoT networks?}

Table \ref{table:Data Analysis Technique} shows the data analysis techniques considered by the researchers in the selected studies. The number of studies that considered using different machine learning algorithms as their threat mitigation technique is high, the data analysis techniques considered by these studies are the precision and recall metrics. These metrics measure the accuracy of predictions made by the machine learning models. The second most considered technique is the packet inspection and the packet analysis. The packet inspection examines the content of a data packet as it passes through a checkpoint within the network, while packet analysis monitors and analyses network traffic at the packet level. The flow rule and the flow rule management are almost the same, the difference is in the dynamism of the rule \cite{42}. Figure~\ref{fig:Dataset Distribution} shows the distribution of the data analysis techniques used for the datasets available in each study. There are also instances where multiple datasets were used in a primary study \cite{S5} as seen in Figure~\ref{fig:The Approach Chart}. The breakdown is seen in Table~\ref{table:Types of Datasets}.

\begin{table}[h!]
\caption{Data Analysis Technique}
\label{table:Data Analysis Technique}
\centering
\begin{tabular}{|p{5.1cm}|p{4.9cm}|c|}
 \hline
 Data Analysis Technique & Study References & Count \\
 \hline
Packet analysis and inspection & \cite{S1, S4, S6, S7, S22, S23, S27, S29, S30, S31, S32} & 11 \\ \hline
Precision and recall metrics & \cite{S2, S5, S7, S8, S10, S11, S12, S13, S14,  S18, S19, S20, S21, S25, S26, S31, S32}	& 17\\ \hline
Flow rules & \cite{S4, S7, S9, S15, S16, S27, S30, S33} & 8\\ \hline
Performance analysis & \cite{S17}	& 1\\ \hline
Hashing and public key infrastructure method & \cite{S24}	& 1\\ \hline
Authentication method & \cite{S28}	& 1\\ \hline
None & \cite{S3} & 1\\ \hline
\end{tabular}
\end{table}

\subsection{RQ3: What kind of SDN datasets are available to address security issues in SDN-based IoT networks?}

The use of IDS/ IPS datasets is designed to support research in SDN. There are publicly available datasets and self-generated/simulated datasets, both are for research purposes. A total of 7 selected studies generated their datasets (simulated/ live datasets). Studies without dataset in their research work are 16 in number.


\begin{table}[h!]
\caption{Types of Datasets}
\label{table:Types of Datasets}
\centering
\begin{tabular}{|l|p{6.2cm}|c|}
 \hline
 Types of Datasets & Study References & Count \\
 \hline
InSDN dataset & \cite{S5} & 1 \\ \hline
CICIDS2018 dataset & \cite{S5} & 1\\ \hline
Kaggle DDoS dataset & \cite{S5} & 1\\ \hline
UNSW-NB15 dataset & \cite{S10, S11, S22, S26} & 4\\ \hline
NSL-KDD dataset	& \cite{S11, S26} & 2\\ \hline
N-BaIoT 2018 dataset & \cite{S12}	& 1\\ \hline
CICIDS2017 dataset & \cite{S13}	& 1\\ \hline
KDD CUP99 dataset & \cite{S19} & 1\\ \hline
UNB-ISCX dataset & \cite{S25} & 1\\ \hline
CICDDoS2019 dataset	& \cite{S32}	& 1\\ \hline
Simulated/live dataset & \cite{S2, S8, S14, S18, S20, S21}	& 6\\ \hline
No Dataset	& \cite{S1, S3, S4, S6, S7, S9, S15, S16, S17, S23, S24, S27, S28, S29, S30, S31, S33} & 17\\ \hline
\end{tabular}
\end{table}

These datasets are valuable resources for research, development, and testing in the fields of networking, cybersecurity, and data analytics. Most of these datasets can be used for various research purposes, including intrusion detection, anomaly detection, network traffic analysis, machine learning model training, and evaluation of security solutions for SDN and IoT environments. Most of the studies that are without datasets considered using either packet analysis and inspection techniques or some other techniques like flow rule, etc.

\subsection{RQ4: What simulation tools and techniques are used for intrusion detection and prevention in SDN-based IoT networks?}

IoT simulators are software tools that replicate IoT device, network, and application behavior in a virtual environment. 

This question aims to find the type of simulation devices and techniques used by the different researchers in the selected studies, the different testbeds, testbed configuration settings, and evaluation methods as considered in the studies. The most common simulator software used by researchers in the selected studies is the mininet simulator. Mininet is a network simulator that builds a virtual network comprising virtual hosts, controllers, switches, and links. Mininet switches are compatible with OpenFlow, enabling flexible custom routing and Software-Defined Networking, and its hosts run standard Linux network software. It is used for research, learning, debugging, prototyping, testing purposes, etc.

Table \ref{table:Simulation Method} shows the simulation devices used in the selected studies and the number of times used. These simulators offer different features, capabilities, and levels of abstraction. The researchers chose the most suitable tools based on their specific requirements and use cases.
 		
\begin{table}[h!]
\caption{Simulation Methods}
\label{table:Simulation Method}
\centering
\begin{tabular}{|p{4.6cm}|p{5.1cm}|c|}
 \hline
Simulation Devices (Software/Hardware) & Study References & Count \\
 \hline
 Intel i7 & \cite{S13, S21, S26} & 3\\
 \hline
Mininet Simulator & \cite{S2, S3, S5, S8, S9, S10, S15, S16, S17, S18, S19, S20, S23, S25, S29, S31, S32}	& 17 \\
\hline
Ryu Controller & \cite{S3, S7, S17, S22, S25, S32}	& 6\\
\hline
ONOS Controller & \cite{S15, S18, S20, S29} & 4\\
\hline
Floodlight Controller & \cite{S8, S19, S23} & 3\\
\hline
POX Controller	& \cite{S2, S31}	& 2\\
\hline
Matlab Simulator & \cite{S6, S11} & 2\\
\hline
MUD Profiles & \cite{S4, S7} & 2\\
\hline
Pyethereum tester tool	& \cite{S17}	& 1\\
\hline
COOJA Simulator	& \cite{S1, S10, S14, S22, S30} & 5\\
\hline
Android and iOS platforms & \cite{S28} &  1 \\
\hline
Hping3	& \cite{S19} & 1\\
\hline
Maxinet	& \cite{S27} & 1\\
\hline
Containernet & \cite{S27}	& 1\\
\hline
Ganache simulator & \cite{S24}	& 1\\
\hline
Ropsten	& \cite{S24} & 1\\
\hline
Guage simulator	& \cite{S7} & 1\\
\hline
None & \cite{S12, S33} & 2\\
 \hline
\end{tabular}
\end{table}

\section{Threats to Validity}
\label{section:threats}
The results of this SMS could have been affected by the following threats:

\subsubsection{Failure to follow the SMS guidelines} To eliminate this threat, we have systematically selected our primary studies in an unbiased manner and according to the SMS guidelines given by Petersen et al. \cite{37}.

\subsubsection{Researcher's bias} To reduce this threat, the two researchers involved in this SMS met periodically and consensually to resolve all discrepancies and disagreements, aiming to minimize errors and eliminate researcher's bias.

\subsubsection{Data extraction inaccuracy} To mitigate this threat, we set up a data extraction table that serves as a guideline that specifies what data should be extracted from the primary studies against each research question.

\subsubsection{Inadequacies in quality assessment} To mitigate this threat, we formulated the quality level scores and clearly defined the quality assessment criteria. Two researchers independently scored the studies.

\subsubsection{Inability to generalize research findings} To mitigate this threat, we performed a systematic selection of primary studies to generalize the research findings. Additionally, we meticulously categorized the algorithms and techniques employed for threat mitigation in those studies, highlighting areas for future research focus.






\section{Conclusion}
\label{section:conclusions}

This Systematic Mapping Study summarizes the Software Defined Network (SDN) controller architectures to enhance security in Internet of Things (IoT) networks. This mapping study contains an analysis of 33 selected studies. We have highlighted the challenges and the areas of concentration where future work should be focused. We have identified and discussed the existing security issues that could pose a threat to IoT networks and the mitigation techniques used by the researchers in the selected primary studies.

While empirical research surrounding the security enhancement of SDN-based IoT networks has taken shape in the direction of the state-of-the-art machine learning model mitigation techniques on SDN-based IoT networks, there remains a lack of comprehensive studies regarding the types of machine learning models and their effective usage that will efficiently mitigate the risks posed by cyber threats in IoT networks.

Considering the dynamism and sophistication of threats in SDN-based IoT networks, future work should focus more on the SDN controller architecture configuration to properly mitigate the threats, or to reduce the risks.


\section*{Acknowledgments}
Adnan Ashraf received a research grant from the Ulla Tuominen Foundation.

\bibliographystyle{plain}
\bibliography{Ref_Items}

\begin{thebibliography}{10}

\bibitem{S19}
Shivaranjani Anbarsu, Arockia Xavier~Annie Rayan, and Vettriselvi Vetrian.
\newblock Software-defined networking for the internet of things: Securing home
  networks using {SDN}.
\newblock In {\em Real-{Time} {Data} {Analytics} for {Large} {Scale} {Sensor}
  {Data}}, pages 215--270. Elsevier, 2020.

\bibitem{44}
Luigi Atzori, Antonio Iera, and Giacomo Morabito.
\newblock The internet of things: A survey.
\newblock {\em Computer Networks}, 54(15):2787--2805, October 2010.

\bibitem{S30}
Jalal Bhayo, Sufian Hameed, and Syed~Attique Shah.
\newblock An efficient counter-based {DDoS} attack detection framework
  leveraging software defined {IoT} ({SD}-{IoT}).
\newblock {\em IEEE Access}, 8:221612--221631, 2020.

\bibitem{S1}
Jalal Bhayo, Riaz Jafaq, Awais Ahmed, Sufian Hameed, and Syed~Attique Shah.
\newblock A time-efficient approach toward {DDoS} attack detection in {IoT}
  network using {SDN}.
\newblock {\em IEEE Internet Things J.}, 9(5):3612--3630, March 2022.

\bibitem{S21}
Jalal Bhayo, Syed~Attique Shah, Sufian Hameed, Awais Ahmed, Jamal Nasir, and
  Dirk Draheim.
\newblock Towards a machine learning-based framework for {DDOS} attack
  detection in software-defined {IoT} ({SD}-{IoT}) networks.
\newblock {\em Engineering Applications of Artificial Intelligence},
  123:106432, August 2023.

\bibitem{S31}
Suman~Sankar Bhunia and Mohan Gurusamy.
\newblock Dynamic attack detection and mitigation in {IoT} using {SDN}.
\newblock In {\em 2017 27th {International} {Telecommunication} {Networks} and
  {Applications} {Conference} ({ITNAC})}, pages 1--6, Melbourne, VIC, November
  2017. IEEE.

\bibitem{S2}
P.K Binu, Deepak Mohan, and E.M Sreerag~Haridas.
\newblock An {SDN}-based prototype for dynamic detection and mitigation of
  {DoS} attacks in {IoT}.
\newblock In {\em 2021 {Third} {International} {Conference} on {Inventive}
  {Research} in {Computing} {Applications} ({ICIRCA})}, pages 5--10,
  Coimbatore, India, September 2021. IEEE.

\bibitem{S29}
Daniele Bringhenti, Jalolliddin Yusupov, Alejandro~Molina Zarca, Fulvio
  Valenza, Riccardo Sisto, Jorge~Bernal Bernabe, and Antonio Skarmeta.
\newblock Automatic, verifiable and optimized policy-based security enforcement
  for {SDN}-aware {IoT} networks.
\newblock {\em Computer Networks}, 213:109123, August 2022.

\bibitem{S32}
Mimi Cherian and Satishkumar~L. Varma.
\newblock Secure {SDN}–{IoT} framework for {DDoS} attack detection using deep
  learning and counter based approach.
\newblock {\em J Netw Syst Manage}, 31(3):54, July 2023.

\bibitem{40}
Tore Dybå and Torgeir Dingsøyr.
\newblock Empirical studies of agile software development: {A} systematic
  review.
\newblock {\em Information and Software Technology}, 50(9-10):833--859, August
  2008.

\bibitem{41}
Tore Dybå, Torgeir Dingsøyr, and Geir~K. Hanssen.
\newblock Applying systematic reviews to diverse study types: An experience
  report.
\newblock In {\em First {International} {Symposium} on {Empirical} {Software}
  {Engineering} and {Measurement} ({ESEM} 2007)}, pages 225--234, Madrid,
  Spain, September 2007. IEEE.

\bibitem{S24}
Zakaria~Abou El~Houda, Abdelhakim Hafid, and Lyes Khoukhi.
\newblock Co-{IoT}: A collaborative {DDoS} mitigation scheme in {IoT}
  environment based on blockchain using {SDN}.
\newblock In {\em 2019 {IEEE} {Global} {Communications} {Conference}
  ({GLOBECOM})}, pages 1--6, Waikoloa, HI, USA, December 2019. IEEE.

\bibitem{S28}
Liming Fang, Yang Li, Xinyu Yun, Zhenyu Wen, Shouling Ji, Weizhi Meng, Zehong
  Cao, and M.~Tanveer.
\newblock {THP}: A novel authentication scheme to prevent multiple attacks in
  {SDN}-based {IoT} network.
\newblock {\em IEEE Internet Things J.}, 7(7):5745--5759, July 2020.

\bibitem{S7}
Ayyoob Hamza, Hassan~Habibi Gharakheili, Theophilus~A. Benson, and Vijay
  Sivaraman.
\newblock Detecting volumetric attacks on {loT} devices via {SDN}-based
  monitoring of {MUD} activity.
\newblock In {\em Proceedings of the 2019 {ACM} {Symposium} on {SDN}
  {Research}}, pages 36--48, San Jose CA USA, April 2019. ACM.

\bibitem{S4}
Ayyoob Hamza, Hassan~Habibi Gharakheili, and Vijay Sivaraman.
\newblock Combining {MUD} policies with {SDN} for {IoT} intrusion detection.
\newblock In {\em Proceedings of the 2018 {Workshop} on {IoT} {Security} and
  {Privacy}}, pages 1--7, Budapest Hungary, August 2018. ACM.

\bibitem{S12}
Tooba Hasan, Adnan Akhunzada, Thanassis Giannetsos, and Jahanzaib Malik.
\newblock Orchestrating {SDN} control plane towards enhanced {IoT} security.
\newblock In {\em 2020 6th {IEEE} {Conference} on {Network} {Softwarization}
  ({NetSoft})}, pages 457--464, Ghent, Belgium, June 2020. IEEE.

\bibitem{S5}
Vanlalruata Hnamte, Ashfaq~Ahmad Najar, Hong Nhung-Nguyen, Jamal Hussain, and
  Manohar~Naik Sugali.
\newblock {DDoS} attack detection and mitigation using deep neural network in
  {SDN} environment.
\newblock {\em Computers \& Security}, 138:103661, March 2024.

\bibitem{S33}
Jiejun Hu, Martin Reed, Nikolaos Thomos, Mays~F. AI-Naday, and Kun Yang.
\newblock Securing {SDN}-controlled {IoT} networks through edge blockchain.
\newblock {\em IEEE Internet Things J.}, 8(4):2102--2115, February 2021.

\bibitem{38}
Samireh Jalali and Claes Wohlin.
\newblock Systematic literature studies: database searches vs. backward
  snowballing.
\newblock In {\em Proceedings of the {ACM}-{IEEE} international symposium on
  {Empirical} software engineering and measurement}, pages 29--38, Lund Sweden,
  September 2012. ACM.

\bibitem{S15}
Kallol~Krishna Karmakar, Vijay Varadharajan, Surya Nepal, and Uday Tupakula.
\newblock {SDN}-enabled secure {IoT} architecture.
\newblock {\em IEEE Internet Things J.}, 8(8):6549--6564, April 2021.

\bibitem{47}
Ajay Kumar, K.~Abhishek, M.R. Ghalib, A.~Shankar, and X.~Cheng.
\newblock Intrusion detection and prevention system for an {IoT} environment.
\newblock {\em Digital Communications and Networks}, 8(4):540--551, August
  2022.

\bibitem{S16}
Cheng Li, Zhengrui Qin, Ed~Novak, and Qun Li.
\newblock Securing {SDN} infrastructure of {IoT–Fog} networks from {MitM}
  attacks.
\newblock {\em IEEE Internet Things J.}, 4(5):1156--1164, October 2017.

\bibitem{S11}
Ke~Luo.
\newblock A distributed {SDN}-based intrusion detection system for {IoT} using
  optimized forests.
\newblock {\em PLoS ONE}, 18(8):e0290694, August 2023.

\bibitem{S20}
Jie Ma, Wei Su, Yikun Li, and Yihua Peng.
\newblock Synchronizing {DDoS} detection and mitigation based graph learning
  with programmable data plane, {SDN}.
\newblock {\em Future Generation Computer Systems}, 154:206--218, May 2024.

\bibitem{S27}
Tri~Gia Nguyen, Trung~V. Phan, Binh~T. Nguyen, Chakchai So-In, Zubair~Ahmed
  Baig, and Surasak Sanguanpong.
\newblock {SeArch}: A collaborative and intelligent {NIDS} architecture for
  {SDN}-based cloud {IoT} networks.
\newblock {\em IEEE Access}, 7:107678--107694, 2019.

\bibitem{37}
Kai Petersen, Sairam Vakkalanka, and Ludwik Kuzniarz.
\newblock Guidelines for conducting systematic mapping studies in software
  engineering: {An} update.
\newblock {\em Information and Software Technology}, 64:1--18, August 2015.

\bibitem{S17}
Shalli Rani, Himanshi Babbar, Gautam Srivastava, Thippa~Reddy Gadekallu, and
  Gaurav Dhiman.
\newblock Security framework for internet-of-things-based software-defined
  networks using blockchain.
\newblock {\em IEEE Internet Things J.}, 10(7):6074--6081, April 2023.

\bibitem{S25}
Nagarathna Ravi and S.~Mercy Shalinie.
\newblock Learning-driven detection and mitigation of {DDoS} attack in {IoT}
  via {SDN}-{Cloud} architecture.
\newblock {\em IEEE Internet Things J.}, 7(4):3559--3570, April 2020.

\bibitem{46}
P.P. Ray.
\newblock A survey on internet of things architectures.
\newblock {\em Journal of King Saud University - Computer and Information
  Sciences}, 30(3):291--319, July 2018.

\bibitem{45}
Karen Rose, Scott Eldridge, and Lyman Chapin.
\newblock The {internet} of {things}: {An} {overview}.
\newblock {\em The internet society (ISOC)}, 80(15):1--53, October 2015.

\bibitem{34}
Farag Sallabi, Faisal Naeem, M.~Awad, and Khaled Shuaib.
\newblock Managing {IoT}-based smart healthcare systems traffic with software
  defined networks.
\newblock 06 2020.

\bibitem{S8}
Abimbola~O. Sangodoyin, Mobayode~O. Akinsolu, Prashant Pillai, and Vic Grout.
\newblock Detection and classification of {DDoS} flooding attacks on
  software-defined networks: A case study for the application of machine
  learning.
\newblock {\em IEEE Access}, 9:122495--122508, 2021.

\bibitem{S18}
Alper~Kaan Sarica and Pelin Angin.
\newblock Explainable security in {SDN}-based {IoT} networks.
\newblock {\em Sensors}, 20(24):7326, December 2020.

\bibitem{S9}
Muhammad~Arslan Sarwar, Majid Hussain, Muhammad~Usman Anwar, and Mudassar
  Ahmad.
\newblock {FlowJustifier}: An optimized trust-based request prioritization
  approach for mitigation of {SDN} controller {DDoS} attacks in the {IoT}
  paradigm.
\newblock In {\em Proceedings of the 3rd {International} {Conference} on
  {Future} {Networks} and {Distributed} {Systems}}, pages 1--9, Paris France,
  July 2019. ACM.

\bibitem{S10}
Qaisar Shafi, Abdul Basit, Saad Qaisar, Abigail Koay, and Ian Welch.
\newblock Fog-assisted {SDN} controlled framework for enduring anomaly
  detection in an {IoT} network.
\newblock {\em IEEE Access}, 6:73713--73723, 2018.

\bibitem{S22}
Qaisar Shafi, Saad Qaisar, and Abdul Basit.
\newblock Software defined machine learning based anomaly detection in fog
  based {IoT} network.
\newblock In Sanjay Misra, Osvaldo Gervasi, Beniamino Murgante, Elena Stankova,
  Vladimir Korkhov, Carmelo Torre, Ana Maria~A.C. Rocha, David Taniar,
  Bernady~O. Apduhan, and Eufemia Tarantino, editors, {\em Computational
  {Science} and {Its} {Applications} – {ICCSA} 2019}, volume 11622, pages
  611--621. Springer International Publishing, Cham, 2019.
\newblock Series Title: Lecture Notes in Computer Science.

\bibitem{43}
Maninder~Pal Singh and Abhinav Bhandari.
\newblock New-flow based {DDoS} attacks in {SDN}: {Taxonomy}, rationales, and
  research challenges.
\newblock {\em Computer Communications}, 154:509--527, March 2020.

\bibitem{S14}
N.~Sivanesan and K.~S. Archana.
\newblock Detecting distributed denial of service ({DDoS}) in {SD}-{IoT}
  environment with enhanced firefly algorithm and convolution neural network.
\newblock {\em Opt Quant Electron}, 55(5):393, May 2023.

\bibitem{S13}
Omer~Elsier Tayfour, Azath Mubarakali, Amira~Elsir Tayfour, Muhammad~Nadzir
  Marsono, Entisar Hassan, and Ashraf~M. Abdelrahman.
\newblock Adapting deep learning-{LSTM} method using optimized dataset in {SDN}
  controller for secure {IoT}.
\newblock {\em Soft Comput}, May 2023.

\bibitem{S26}
Qiuting Tian, Dezhi Han, Meng-Yen Hsieh, Kuan-Ching Li, and Arcangelo
  Castiglione.
\newblock A two-stage intrusion detection approach for software-defined {IoT}
  networks.
\newblock {\em Soft Comput}, 25(16):10935--10951, August 2021.

\bibitem{39}
Muhammad Usman, Emilia Mendes, Francila Weidt, and Ricardo Britto.
\newblock Effort estimation in agile software development: a systematic
  literature review.
\newblock In {\em Proceedings of the 10th {International} {Conference} on
  {Predictive} {Models} in {Software} {Engineering}}, pages 82--91, Turin
  Italy, September 2014. ACM.

\bibitem{42}
Qing Wei, David Perez-Caparros, and Artur Hecker.
\newblock Dynamic flow rules in software defined networks.
\newblock In {\em 2016 {Fifth} {European} {Workshop} on {Software}-{Defined}
  {Networks} ({EWSDN})}, pages 25--30, Den Haag, Netherlands, October 2016.
  IEEE.

\bibitem{S6}
Tong Xu, Deyun Gao, Ping Dong, Hongke Zhang, Chuan~Heng Foh, and Han-Chieh
  Chao.
\newblock Defending against new-flow attack in {SDN}-based internet of things.
\newblock {\em IEEE Access}, 5:3431--3443, 2017.

\bibitem{S23}
Da~Yin, Lianming Zhang, and Kun Yang.
\newblock A {DDoS} attack detection and mitigation with software-defined
  internet of things framework.
\newblock {\em IEEE Access}, 6:24694--24705, 2018.

\bibitem{S3}
Yuyang Zhou, Guang Cheng, and Shui Yu.
\newblock An {SDN}-enabled proactive defense framework for {DDoS} mitigation in
  {IoT} networks.
\newblock {\em IEEE Trans.Inform.Forensic Secur.}, 16:5366--5380, 2021.

\end{thebibliography}

\end{document}